# THE LMC SUPERSOFT X-RAY BINARY RX J0513.9−6951

David Crampton[1] and J.B. Hutchings[1]

Dominion Astrophysical Observatory, National Research Council of Canada,
Victoria, B.C. V8X 4M6, Canada

A.P. Cowley[1], P.C. Schmidtke[1], and T.K. McGrath[1]

Department of Physics and Astronomy, Arizona State University, Tempe, AZ, 85287

D. O'Donoghue and M.K. Harrop-Allin

Department of Astronomy, University of Cape Town, Rondebosch 7700, South Africa

## ABSTRACT

A detailed analysis of simultaneous photometric and spectroscopic observations of the optical counterpart of the LMC "supersoft" X-ray source RX J0513.9−6951 (identified with HV 5682) is presented. The spectrum is dominated by He II emission lines and H + He II blends; no He I is observed but several higher ionization emission features, especially O VI (3811, 3834, and 5290Å) are prominent. Radial velocity measurements suggest a binary period of $0\overset{d}{.}76$. If the small velocity amplitude, K∼11 km s$^{-1}$, is interpreted as orbital motion, this implies that the binary system contains a somewhat evolved star plus a relatively massive compact object, viewed nearly pole-on. No orbital photometric variations were found, although irregular brightness changes of ∼ 0.3 mag occurred. Unusual emission lines are observed which cannot be identified except as high velocity (4000 km s$^{-1}$) bipolar outflows or jets. These outflows are seen in H and He II at the same positive and negative velocities. They were relatively stable for periods of ∼5 days, but their velocities appear to have been ∼250 km s$^{-1}$ smaller in 1992 than in 1993 or 1994.

*Subject headings:* accretion disks – stars: individual (RX J0513.9−6951) – X-rays: stars

## 1. INTRODUCTION

The "supersoft" transient X-ray source RX J0513.9-6951 (hereafter referred to as 0513−69) was discovered in outburst during the ROSAT All Sky Survey by Schaeidt, Hasinger, and Trumper

---

(1993). It was subsequently identified by Pakull et al. (1993) and Cowley et al. (1993) with a ~16.5 mag peculiar emission-line star in the Large Magellanic Cloud which Cowley et al. (1993) recognized to be Harvard Variable 5682. "Supersoft sources" (SSS) are very luminous X-ray objects (bolometric luminosities of $\sim 10^{38}$ ergs s$^{-1}$) with characteristic blackbody temperatures of kT $\sim$ 30 – 60 eV (see review by Hasinger 1994). They are easily detected in the LMC due to the very low column density of interstellar gas along the line of sight. In many ways 0513−69 is similar to CAL 83 (Long, Helfand, and Grabelsky 1981), the prototype SSS, which is a well-known LMC X-ray binary with an orbital period of ~1 day. Recently, extensive observations of a SSS in the Galaxy, RX J0019.8+2156 (hereafter referred to as 0019+21), have been reported (Beuermann et al. 1995; Greiner and Wenzel 1995). Fortunately, this galactic source is sufficiently bright that its photometric history can be followed for the past century from archival photographic plates. Most of its characteristics are very similar to SSS in the LMC, and it can be compared to 0513−69.

Although the SSS have now been studied for several years, there is still no consensus as to the nature of the underlying systems. Virtually all models require accretion near, or even above, the Eddington limit, but whether the compact objects are white dwarfs, neutron stars or black holes is still being debated. The high X-ray luminosities and optically bright accretion disks suggest very high mass transfer/accretion rates, well above that in traditional low-mass X-ray binary systems (LMXB) like Sco X-1 or LMC X-2. Broad emission and P Cygni absorption profiles observed in the spectra of several SSS are additional evidence of high mass transfer. Currently, the most popular model involves steady nuclear burning on the surface of an accreting white dwarf (e.g., van den Heuvel et al. 1992; Pakull et al. 1993), but mass estimates from dynamical studies of these binary systems favor higher mass compact objects rather than white dwarfs, as is described below. If the compact objects are neutron stars, the softness of the X-ray spectra indicates they must be surrounded by some type of cocoon which down-scatters the X-rays (Greiner, Hasinger, and Kahabka 1991; Kylafis and Xilouris 1993).

Although 0513−69 appears to be an almost proto-typical SSS, it displays some very unusual characteristics. During the X-ray outburst in 1990 the optical brightness may have marginally declined (Pakull et al. 1993) rather than increasing as most transient X-ray sources do. It is also the first SSS to exhibit repeated X-ray outbursts on timescales of years (Hasinger 1994). The spectrum is characterized by the extreme strength of the He II 4686Å line emission consisting of 4000 km s$^{-1}$ wide wings, a peak flux approximately three times the continuum, and an equivalent width of ~20Å. Highly ionized emission lines of C, N and O are also present. Most unusual are red- and blue-shifted emission components of the strongest H and He II lines, with relative radial velocities of $\sim \pm 4000$ km s$^{-1}$ (Cowley et al. 1995).

Our preliminary observations of 0513−69 in 1992, as well as those of Pakull et al. (1993), indicated that small radial velocity variations are present, so further photometric and spectroscopic observations were carried out in 1993 and 1994 to try to establish the parameters of the binary system.

## 2. OBSERVATIONS AND MEASUREMENTS

### 2.1. Photometry

The optical counterpart of 0513−69 has been known to be variable for more than 100 years. It was identified as Harvard Variable 5682 by Leavitt (1908) who reported variations of ∼1 mag between 1893 and 1906. The system appears to have shown much smaller variations in recent years. Photometry obtained in 1992 and 1993 (Cowley et al. 1995) shows a range of $V$ magnitudes of only ∼0.3 mag.

To search for orbital photometric variations, a campaign was mounted in November 1994 to obtain observations from both Chile and South Africa over a period of several days. Photometry was obtained with the Tek2048#2 CCD on the CTIO 0.9-m telescope and with a Wright Instruments CCD on the 1.0-m SAAO telescope at the Sutherland observing station. All exposures were through $V$ filters and had integration times between 450 and 500 s. A total of 24 images were taken at CTIO during three consecutive nights, 1994 November 2/3 – 4/5 (UT), typically with ∼$1''\!.4$ seeing. At Sutherland 41 images were acquired, but only on two nights (November 3/4 and 4/5) due to unfavorable weather. Unfortunately, the seeing was particularly poor, $2''\!.4 - 4''\!.2$. All data were reduced using DAOPHOT (Stetson 1987), using differential techniques (Schmidtke 1988), but the SAAO data required special handling as noted below.

The image of 0513−69 (star '1') is blended with an optical companion that lies $1''\!.3$ away at position angle 118° and is ∼2.3 mag fainter in $V$ (see finding chart given by Cowley et al. 1993). This companion contributes about 10% of the blended light. It is resolved on nearly all images from CTIO, but on almost none from Sutherland. To improve the S/N of results from the latter dataset, those images were combined into five groups of eight observations (one image was not used). Within each group, the images were shifted to a common coordinate system, averaged to remove cosmic rays, and then processed using DAOPHOT. Examination of the magnitudes for 0513−69 and its optical companion showed they are highly correlated if both components were left as free parameters (on those images where the components are resolved), so we forced the companion to have $V = 19.048$ on all images and corrected the 0513−69 magnitudes accordingly.

To test this deconvolution procedure, the same technique was applied to star '2' (Cowley et al. 1993), which is a slightly wider pair (separation $3''\!.8$) with a much smaller magnitude difference ($\Delta V \sim 0.6$). Using only CTIO data, the magnitude of the brighter component of star '2' is $V = 16.635\pm0.012$. For comparison, the SAAO images yield $V = 16.550\pm0.054$. Hence, the two datasets for this test case differ by 0.085 mag, or $1.3\sigma$.

Table 1 and Figure 1 summarize the 1994 photometric data for 0513−69. The single point near HJD2449660.4 comes from a group of eight Sutherland images and is slightly brighter than the CTIO magnitudes. The brightening is consistent with that found for the test case (star '2') and is not considered significant. On the other hand, the SAAO data points near HJD2449661.5 are considerably fainter than all of the CTIO photometry. Since the optical companion contributes just 10% of the total light, even a 100% error in compensating for its presence could only account for 0.1 mag error in 0513−69. Hence, the 0.3 mag dimming appears to be real. The 1994 CTIO photometry yields $V = 16.837 \pm 0.012$ when the source is bright. The standard deviation of these measurements (24 in total from three nights) is identical to that found for the test case (star '2').

Using all of the data from 1992, 1993, and 1994, no stable light curve can be found by folding the data from each observing run on trial periods from $0\overset{d}{.}2$ to 10 days. When the photometry is folded on the spectroscopic period of $0\overset{d}{.}76$ (discussed in §3), bright and faint phases overlap, and the time of minimum light occurs near a quadrature. None of the aliases of the spectroscopic period shows a smoothly modulated light curve either. Therefore, orbital brightness modulations, if present, must be very small. This suggests that the orbital inclination of the system is very low, a result which is also indicated by the spectroscopic data.

Although the observed brightness changes in 0513−69 appear not to be related to its orbital period, the source does show brighter and fainter 'states', with $V \sim 16.7 - 16.8$ when it is 'bright' and $V \sim 17.0 - 17.1$ when it is 'faint'. The division is not as clearly defined in the $B$ filter measurements of Pakull et al. (1993), near the time of an X-ray outburst. The photometric behavior is similar to that observed for 0019+21 (Greiner and Wenzel 1995), which shows irregular fluctuations (up to 0.5 mag) on timescales of weeks to months. In 0513−69 the low states can be quite brief; the minimum observed in November 1994 lasted less than a day.

We further note that scattered data indicate that CAL 83, which shows orbital modulations of $\Delta V \sim 0.25$ mag, also shows long term trends in its light curve. Smale et al. (1988) found the mean $V \sim 16.3$ in August 1984, while in December 1984 the overall brightness of the system had declined to a mean $V \sim 16.9$. Our photometry of CAL 83 obtained in November 1985 showed the average $V$ magnitude had dropped to 17.3 (Crampton et al. 1987). We have no further photometry until Dec. 1993 when we found CAL 83 at $V \sim 17.5$ in a single observation. Thus, there seems to be a long term change in its average brightness. These non-orbital variations seen in the SSS are probably due to changes in the overall disk brightness, but at present there is no clear model to explain them.

A magnitude of $V \sim 16.8$ for 0513−69 corresponds to $M_V \sim -2$, assuming a LMC distance modulus of 18.5 (van den Bergh 1992) and average reddening $E_{B-V} = 0.1$ mag. Since the $V$ light appears to be completely dominated by emission from an accretion disk, the luminosity of the disk is among the brightest known, being comparable to the prototype SSS, CAL 83. For comparison, among 18 LMXB discussed by van Paradijs and McClintock (1994), only the X-ray nova V404 Cyg is brighter. Most disk dominated systems are fainter than $M_V \sim 0$.

## 2.2. Spectroscopy

### 2.2.1. Data

The spectroscopic data were obtained with the CTIO 4-m telescope, using a long slit on the RC spectrograph in 1992 and 1994 and the ARGUS multifiber system in 1993. The 1992 December spectra cover the wavelength range 3875–5030Å with a resolution of ∼0.9Å per pixel. The ARGUS spectra include the range 3650–5800Å with a resolution of ∼1.8Å per pixel. In 1994, two gratings were used, giving ∼0.9Å per pixel in the 3900–5020Å region and ∼1.8Å per pixel in the 3720–5850Å region, respectively.

One-dimensional spectra were extracted and processed following standard techniques with IRAF to yield wavelength-calibrated spectra. Calibration data were usually taken before and after each stellar exposure, and the wavelengths are established to ∼0.1 pixel or better. In view of the narrow slit used and variable seeing conditions, no flux calibrations were attempted. Details of the spectroscopic measurements are given in Tables 2, 3, and 4.

### 2.2.2. The Spectrum of 0513−69

The average spectrum from all our 1994 low resolution spectra of 0513−69 is shown in Figure 2, with a single spectrum of CAL 83 taken with the same instrument for comparison. As Figure 2 demonstrates, the spectrum is dominated by He II emission lines and H + He II blends. No He I is observed, but several higher ionization emission features, especially O VI (3811, 3834, and 5290Å) are prominent in both 0513−69 and CAL 83. There is probably a weak emission line of O V near 5595Å, and N V is represented by emission lines at 4603 and 4619Å. Weak emissions due to C IV 5801Å and 5812Å and possibly 4658Å (blended with the violet wing of He II) appear to be present, although we note that C IV 1549Å was not observed in the UV by Pakull et al. (1993). The situation with C III is confused: 5695Å emission is not present, but lines near 4650Å from another C III multiplet appear to contribute to the blue wing of He II 4686 (see §4.1), and a line near 4070Å is also likely to be due to C III. Hence, in spite of the report by Pakull et al. (1993) that carbon may be deficient, there is evidence for carbon lines in our spectra. Average equivalent widths of the principal emission features are listed in Table 3. Also tabulated are data on features which we believe to be doppler-shifted emission components of He II 4686Å and Hβ. These features are discussed in §4.2.

### 2.2.3. Spectral Variations

The 1993 spectra differ from the 1992 and 1994 spectra in several respects. The most obvious change is that in 1993 the He II emission lines were ∼50% weaker, the H lines were ∼40% weaker, and the line profiles were different. These variations in the strength and profiles of Hβ and He II 4686Å are very apparent in Figure 3. The red wing of the emission is much broader in 1994 than 1993. Note that the extended violet wing of He II 4686Å does not appear to have changed intensity, indicating that either the broad wing itself or the other contributing lines did not vary as much as the sharp components of H and He II. The variations in line strengths were not accompanied by large changes in the overall brightness of the system – the mean V magnitude was 16.78 in 1992 December, 16.99 in 1993 and 16.84 in 1994. No significant changes in line strengths or profiles were observed during individual observing runs (i.e., on timescales of ~5 days) apart from small changes in the "shifted lines" described below.

### 2.2.4. Spatially Extended Line Emission?

Pakull and Angebault (1986) and Pakull and Motch (1989) first reported detection of resolved ionized nebulae surrounding X-ray sources, including one around the SSS CAL 83. More recently, Rappaport et al. (1994) developed detailed models for ionization nebulae expected around SSS. Subsequently Remillard, Rappaport, and Macri (1995) obtained [OIII] and H$\alpha$ images to search for such nebulae in the Magellanic Clouds. Although they report successful observations of the nebula around CAL 83, no nebulosity was detected around 0513−69 or any of the other eight SSS. Our long slit observations of 0513−69 similarly show no evidence of spatially extended line emission at [OIII], H$\beta$ or He II, although comparable observations of CAL 83 show extended emission at all of these lines.

## 3. THE 0513−69 BINARY SYSTEM

### 3.1. Radial Velocity Measurements

Radial velocities were measured from the spectra in several ways. First, the strong line peaks were measured individually by fitting parabolae through them. The velocity variation is very small, amounting to only ~20% of a pixel in the lower resolution data. The emission lines are also quite narrow, ~6Å, and hence are only sampled by a few pixels in the lower resolution data. Thus, it is difficult to measure the line centers to the required accuracy. Another complication is that H lines are blended with He II so, for example, the velocities of H$\beta$ systematically differ by −17 km s$^{-1}$ from those of the He II lines. Nonetheless, the velocities of He II 4686, He II 4541, H$\beta$, and H$\gamma$ were measured individually on all spectra, and all show similar velocity variations.

Second, individual spectra were cross correlated against a mean spectrum for each year and spectrographic configuration. Since the line strengths and profiles vary and the datasets have different sampling and signal-to-noise in various spectral regions, it is not straightforward to tie the cross-correlation velocities together reliably. Within a dataset, however, the observed velocity variations from cross correlation generally agree very closely with those for the He II peaks measured individually, particularly for the best data (the 1994 higher dispersion spectra). Attempts were made to measure the velocites of only the weaker lines by cross correlation, restricting the wavelength region to 4100−4600Å. This worked reasonably well in the higher resolution 1994 data, but for the other spectra this wavelength region is too noisy to give reliable results. The velocity amplitude might be somewhat higher for these weaker lines, but since the velocity minimum is not covered with the higher dispersion observations, the result is inconclusive.

Ultimately, we concluded that the most accurate velocities were obtained by cross correlating the whole spectrum against the mean for each run and then correcting each dataset to bring them to a common zeropoint. We note that this cannot be done with confidence due to the differences in wavelength region, spectral resolution, and changing line profiles in the various datasets. However, the velocity variations observed during our three runs all showed very similar amplitudes to each other and to the combined data. The velocities corresponding to the peaks of He II 4686Å and to the cross correlation are given in Table 2 for all our data. The 1994 velocities are plotted in Figure 4, demonstrating the good agreement between the He II 4686Å velocities and those derived from cross correlation of the whole spectrum.

### 3.2. Period and Orbit

The velocity variations observed in the 1993 dataset indicated a possible binary period of P = $0^d.43$, with alias periods of $0^d.32$ and $0^d.77$ (Cowley et al. 1995). The 1994 data show similar velocity variations, but it is still difficult to determine which period is best, even with the combined data from three years. The longer period, near $0^d.76$, formally gives the highest velocity amplitude and smallest residuals, but periods near $0^d.43$ cannot be ruled out. Aliases at even longer periods, for example at $\sim 1^d.2$ and $4^d.8$, produce residuals which have strong hour angle dependences indicating that they are most likely spurious. Although the 'best' period is $0^d.76$, the exact value is indeterminate with possible periods differing by 0.0017 days, an interval which corresponds to one cycle difference between the 1993 and 1994 observing runs. The combined velocity curve differs only slightly between various periods near $0^d.76$. The radial velocities are shown in Figure 5, folded on the ephemeris T = HJD 2449332.63 + 0.75952E, where T is the epoch of maximum positive velocity.

The velocities from either the He II peak or from cross correlation of the whole spectrum are best fit with circular orbits of velocity semi-amplitude, K $\sim 11 \pm 2$ km s$^{-1}$ (see Figure 5). The amplitude is so small and the velocities sufficiently inaccurate that formal fits to non-circular orbits are not meaningful. The short period and low velocity amplitude imply a very low mass function, f(M) = $0.00011 M_\odot$. If the emission lines are produced on or near the compact object and represent orbital motion, then strong conclusions can be drawn from such a small mass function. Figure 6 shows how the masses of the component stars are related for various angles of orbital inclination, $i$. If the compact object is a white dwarf with $0.4 < M_x < 1.2 M_\odot$, then the mass of the companion $M_c < 0.2 M_\odot$ for inclinations greater than $i \sim 15°$. It is very unlikely that such a low mass star could have evolved to fill its Roche lobe, unless it has lost considerable mass since it left the main sequence. If the mass of the non-degenerate companion is increased to a point where a main sequence star could reasonably fill its Roche lobe then the inclination must still be very low and $M_x >$ a few solar masses. In the latter case, Figure 6 shows that unless the inclination is $\leq 20°$, the mass of the compact object becomes very large. Although the probability of viewing the system at such a restricted inclination angle would normally be small, it is not inconceivable, since 0513−69 has been pre-selected by the strength of its soft X-ray emission which may favor pole-on systems. Nonetheless, Figure 6 shows that if the measured value of K represents orbital motion, the compact object is likely to be more massive than a white dwarf, unless the mass-losing star has an extremely small mass ($<0.2 M\odot$). Since the mass function is very sensitive to the value of velocity semi-amplitude, K (f(M) varies as $K^3$), we investigated whether other lines might give larger amplitudes, but no good evidence was found. On the basis of our observations it appears unlikely that K could be more than $\sim 1.5\times$ our measured value, which would not substantially change these conclusions.

Since the mass transfer rate must be very high and variable (as indicated by the erratic X-ray and optical light variations), the companion presumably fills its Roche lobe. As Figure 6 shows, if the compact object is a $\sim 1$ $M_\odot$ white dwarf, the mass of the companion must be $<<1$ $M_\odot$ for any realistic value of the inclination. However, a main-sequence star in this range of masses is far from filling its Roche lobe. A main sequence star would fill its Roche lobe only when $M_c >2M_\odot$ (and $M_x >5M_\odot$) even at $i = 5°$. In other words, a main sequence star would not fill its Roche lobe anywhere in Figure 6. Thus, the companion must be evolved. However, for that to be the case the mass cannot be much less than $\sim 0.8$ $M_\odot$ or the star would not have had time to evolve, unless extreme mass loss has occurred. On the other hand, $M_c$ cannot be too massive or it would have evolved to the point where its luminosity is so high as to be detectable either by red colors or by the presence of late-type spectral features (e.g., Mg b). For reasonable values of the companion mass, $\sim 1$ $M_\odot$, the mass of the compact object appears to be much greater than a white dwarf, unless we happen to be viewing the system exactly pole-on.

### 3.3. Phase-binned Spectra

Spectra with comparable signal were co-added in bins near the phases of maximum and minimum radial velocity and at the quadratures to look for changes that might be related to the binary system structure. There is very little difference among the groups, even in our best (1994) data. The only marginal changes seen are $<10\%$ in total flux and only $\sim 2\sigma$ in significance. The H$\beta$ wind absorption appears to be strongest at phase 0.75, when the X-ray source with its bright disk is closest to the observer and its wind is directed away from the companion. The higher ionization lines are weakest at phase 0, when the disk is receding at maximum velocity. This is also when the redward-shifted high velocity component of H$\beta$ is weakest. We caution that these changes are of low significance and may not be phase related.

### 4. HIGH VELOCITY OUTFLOWS

### 4.1. Broad wings

In the spectra of 0513−69 and CAL 83, broad high-velocity wings of the strongest emission lines are present, particularly at He II 4686Å. In CAL 83, observations over several years demonstrate that the emission wing is sometimes on the violet side of the main peak and sometimes on the red side (Crampton et al. 1987). In all our 0513−69 spectra the strong 4686Å emission line shows a broad wing which extends $\sim 60$Å, or $\sim 4000$ km s$^{-1}$, to the violet. On the other hand, H$\beta$ shows evidence of wind absorption over a similar velocity range, $\sim -300$ km s$^{-1}$ to the shifted line at $\sim -4000$ km s$^{-1}$.

The redward wings of H$\beta$ and He II 4686Å were used to construct a smooth wing profile for the 1994 data, which was then reflected about the center of the line to measure the asymmetrical wings. Figure 7 shows the broad wing of He II 4686Å, with the location of possible emission lines indicated. But, as described in §2.2.2, it is not clear whether all of these lines actually contribute to the overall emission profile. Based on the strengths of other lines of C III, C IV, and N III, it appears likely that there is considerable underlying He II 4686Å emission. It is perhaps notable that the emission and absorption features exhibit similar velocity ranges, up to nearly 4000 km s$^{-1}$. The broad wings may arise in a different region from the sharp peak, but our attempts to measure possible variations in their radial velocity have produced only upper limits of order 100 km s$^{-1}$. In view of the observed violet-displaced absorption of H$\beta$, it is surprising that the IUE spectrum obtained by Pakull et al. (1993) shows little or no such absorption. The absence of the ultraviolet emission lines of C IV and Si IV are also puzzling. Further UV spectra should be obtained.

### 4.2. Bipolar Outflows?

Extraordinary emission lines located on either side of the strongest emission features in the spectrum of 0513−69 were tentatively identified as high velocity components of He II 4686Å and H$\beta$ independently by Cowley et al. (1995) and Pakull (1994, private communication). Our November 1994 spectra with higher signal-to-noise and better resolution strengthen this identification. Even though an exhaustive search (see below) was carried out for any plausible identifications of these features with ions which might be anticipated (including very high ionization lines), none were found. However, the features display a constant delta-lambda over lambda relationship with the He II 4686Å and H$\beta$ lines, strongly suggesting a doppler origin.

Average equivalent widths and wavelengths of these displaced lines for each year are given in Tables 3 and 4. Two values are given for the 1994 data, corresponding to the low and high resolution spectra. Average spectra (showing these components) from our three observing seasons are displayed in Figure 8, with the features marked S$^+$ and S$^-$. Even though the 1992 average spectrum is of lower quality, the lines apparently have not changed very much during the 1992

– 1994 period. If the average 1993 and 1994 wavelengths listed in Table 4 are corrected for a systemic velocity of +280 km s$^{-1}$ (for 4686Å and +263 km s$^{-1}$ for H$\beta$ to allow for He II), the restframe wavelengths of the four features are 4623.6Å, 4748.1Å, 4795.5Å, and 4927.1Å. The FWHM of these lines are ∼6Å, except for the latter which is ∼10Å. Since the 4927Å feature in 0513−69 is wider than the other observed features, it is likely that it is a blend. In that regard we note that the CAL 83 spectrum shown in Figure 2 has an unidentified emission at ∼4930Å which may contribute to the 4927Å feature in 0513−69, although it is absent in a higher resolution spectrum of CAL 83. A predicted N V line at 4933Å is unlikely to contribute to this features since other expected multiplets of N V are not seen. While there could be a minor contribution of N V 4619Å to the 4624Å feature, similar in strength to the weak N V 4603Å emission line, we know of no other likely lines that would contribute to this feature. Similarly, only some predicted lines (i.e., not observed in the laboratory) of N V and O V, which are expected to be very weak if present at all, could contribute to the other features. We conclude that there are no likely lines which can explain these four features apart from doppler-shifted components of He II and H$\beta$.

If these four emission features are blueward- and redward-shifted components of He II 4686Å and H$\beta$, their relative velocities are shown in the bottom line of Table 4. All the 1993 and 1994 measures are consistent with an average velocity of 4020 km s$^{-1}$. The 1992 spectrum is much noisier, and the velocity appears to be ∼250 km s$^{-1}$ lower (in absolute terms). Possible shifted emission components of H$\gamma$ are also present in our 1994 spectra at 4404Å and perhaps at 4283Å. The average equivalent widths of the shifted lines are all similar, ∼0.7Å, with the exception of the violet component of H$\beta$. The measurement of the latter feature is uncertain since it is in a region affected by the broad wind absorption, making the placement of the continuum difficult.

Although most emission lines of 0513−69 were ∼50% weaker in 1993, the strength of these shifted features appears to have remained relatively constant (see Table 3) during the 1992–4 period. During our observing runs, small flux changes were apparent, and some of these displaced components were not seen at all for one or two nights (and possibly longer). For example, on night 5 of the 1994 observing run, both shifted emissions of He II became extended to lower (absolute) velocities by ∼1–2Å. This was the last night of the observing, so we do not know if this was the start of a systematic move or a transient event. In general, however, these four features appear to be stable, sharp emission lines displaced ∼4000 km s$^{-1}$ on either side of He II 4686Å and H$\beta$. We suggest that they are high velocity components of these lines. The fact that these displaced features all moved closer to the parent feature (i.e., to lower velocity) in 1992 supports this interpretation. We conclude that these unique features must be the result of some type of bipolar outflow.

### 4.3. Outflows in CAL 83

Since 0513−69 and CAL 83 are both supersoft sources with similar strong emission-line spectra, two spectra of CAL 83 taken in 1994 Dec were examined for the presence of shifted emission features. Even though we have considerably less data with somewhat lower signal, lines of the same equivalent width as in 0513−69 should be detectable. Nothing is seen, with the possible exception of the weak feature at 4930Å, as discussed above.

As shown by Crampton et al. (1987; Fig. 2), the strengths and shapes of the emission lines in CAL 83 are quite variable on timescales of several months. Sometimes the hydrogen emission disappears completely. The broad wing of He II 4686Å is alternately red- and blue-shifted. Crampton et al. postulated that the changes of this high velocity ($\sim 1000$ km s$^{-1}$) component of He II could be due to precession of the accretion disk and found that period of 69 days would fit all the available data. In November 1994 the broad wings were on the violet side of the line, similar to the profile seen in 1982. This new observation fits the previously-suggested 69-day period, but since so many cycles have passed and the periodicity was so poorly established it isn't possible to constrain it with this single observation. An alternate explanation of the observed variation in CAL 83 is that both broad emission wings and variable broad absorption are present which combine to produce the observed profiles. It would be interesting to monitor the spectra of CAL 83 and 0513−69 for several years to understand better their long-term changes.

## 5. DISCUSSION

The small orbital velocity variation observed in 0513−69 makes it the second SSS with an apparently low velocity amplitude and period near one day, the other being CAL 83. The small semi-amplitude, K, and lack of periodic light variations suggest the binary system is viewed nearly pole-on. The brightness of the accretion disk and lack of soft X-ray absorption lend further support to such a model. The high disk luminosity and X-ray flux indicate very rapid accretion near the Eddington limit. Assuming that the observed velocity variations represent orbital motion, the mass diagram (Figure 6) indicates the compact object's mass is several solar masses for reasonable values of the companion's mass at virtually any inclination angle.

CAL 83 has a period of 1.0475 days (Cowley et al. 1991) and a measured velocity semi-amplitude from the He II 4686Å line of $\sim$35 km s$^{-1}$ (Crampton et al. 1987; Smale et al. 1988). In this case too, the system must be viewed at a low inclination angle and contain a relatively high mass compact object with an evolved companion, if the He II emission velocities represent true orbital motion of the compact star. Analysis of the relative phasing of the light curve, emission-line strengths, and velocity variations in the eclipsing SSS binary CAL 87 demonstrates that He II must be formed near the compact object. CAL 87 has a period of $0\overset{d}{.}44$ and a He II velocity semi-amplitude of K$\sim$40 km s$^{-1}$. The presence of an eclipse indicates the orbital inclination is $> 70°$(Cowley et al. 1990; Pakull et al. 1988). Based on these parameters, Cowley et al. (1990) concluded that CAL 87 contains an evolved, mass-losing star of mass $> 0.5 M_\odot$ and a black hole of mass $>6 M_\odot$.

Recently the galactic SSS 0019+21 has been discussed by Beuermann et al. (1995). They have found an orbital period of P = $0^{d}\!.66$ and measured the velocity semi-amplitude to be K = 67 km s$^{-1}$. The modulation of the X-ray and optical light curves indicates that the system must have a moderate inclination, although it is not eclipsing. These parameters again imply the most probable masses for the compact object are larger than white dwarf masses, even though the authors discuss the system in terms of a white dwarf model. Thus, it seems possible that the four systems are fairly similar, with CAL 83 and 0513−69 being viewed at very low inclination angles while 0019+21 and CAL 87 are seen at higher angles. The dynamical evidence suggests the underlying stars are very similar. In all these cases, the companion stars must be evolved in order to fill their Roche lobes. Hence, their masses cannot be extremely small, unless they have lost a significant fraction of their mass during their earlier evolution. For values of the companion star masses in the range ∼0.7–1.0M$_\odot$, the masses of the compact objects appear to be greater than the Chandrasekhar limit for a white dwarf.

One characteristic of 0513−69 which is difficult to explain is the fact that the optical luminosity probably did not change during the observed X-ray outburst in 1990. With a white dwarf model, Pakull et al. (1993) suggested that the X-ray outburst was triggered by a reduced rate of mass transfer in a relatively high-mass white dwarf system in which accretion is occuring very close to the Eddington limit. According to Kato (1985), a sudden decrease in the radius and corresponding increase in the temperature ensues, giving rise to an X-ray outburst while the optical light from the accretion disk is largely unaffected. Some other scenario would have to be invoked for a neutron star or black-hole model, if this behavior is verified.

Both 0513−69 and CAL 83 show evidence of high velocity flows. The broad violet-displaced emission of He II 4686Å and the P Cygni-type absorption profile of H$\beta$ reach velocities of ∼4000 km s$^{-1}$. In the spectrum of 0513−69, four emission lines are observed which appear to be red-shifted and violet-shifted components of He II 4686Å and H$\beta$, since they maintain a constant delta lambda over lambda relationship with the parent line. These features remained approximately unchanged during our 1992–4 observations. These relatively narrow, slightly variable, "shifted lines" are likely to be the result of bipolar outflows. Two other X-ray binaries are known to have collimated outflows or jets: Cyg X-3 (Strom et al. 1989) and, of course, SS433 (e.g., Margon 1984; Vermuelen 1993). In SS433, the shifted lines have mildly relativistic velocities and are believed to arise in twin jets composed of a series of "bullets", most probably accelerated by the luminosity of an exceptionally bright accretion disk. The dramatic ∼164 day precession of the shifted emission lines in SS433 is presumably caused by spin misalignment in the system rather than properties of the disk. The shifted lines in 0513−69 have much lower velocities, but there is evidence of bullet-like behavior (variable intensity, narrow features which are sometimes double). The 0513−69 system is noteworthy in having a brighter accretion disk (M$_V$ ∼ −2; see §2.1) than most low mass X-ray binaries, and while it is much less luminous than the high mass X-ray binary SS433 (M$_V$ ∼ −6), both systems appear to be accreting near or above the Eddington limit. In summary, the shifted emission lines in the 0513−69 system appear to be evidence of outflows from a luminous accretion disk.

More spectroscopic data are required to confirm and refine the orbital parameters of 0513−69, to prove that the unidentified emission lines are doppler-shifted features and to study their behavior. Further coordinated X-ray and optical monitoring of 0513−69 should be undertaken to confirm that the optical luminosity does not change during X-ray outbursts.

APC and PCS acknowledge support from NSF for this work. We especially thank the staff of CTIO for their assistance with the observations. McGrath's observing expenses were partially supported by a grant from Sigma Xi.

Table 1: Photometry of RX J0513.9−6951 during November 1994

| HJD (2440000+) | V | Observatory | HJD (2440000+) | V | Observatory |
|---|---|---|---|---|---|
| 9659.768 | 16.850 | CTIO | 9659.866 | 16.829 | CTIO |
| 9659.779 | 16.857 | CTIO | 9660.418 | 16.75 | SAAO[a] |
| 9659.785 | 16.857 | CTIO | 9660.799 | 16.869 | CTIO |
| 9659.792 | 16.848 | CTIO | 9660.810 | 16.830 | CTIO |
| 9659.798 | 16.844 | CTIO | 9660.818 | 16.835 | CTIO |
| 9659.804 | 16.847 | CTIO | 9660.825 | 16.825 | CTIO |
| 9659.811 | 16.837 | CTIO | 9660.832 | 16.829 | CTIO |
| 9659.817 | 16.822 | CTIO | 9660.838 | 16.837 | CTIO |
| 9659.823 | 16.837 | CTIO | 9661.426 | 17.08 | SAAO[a] |
| 9659.829 | 16.817 | CTIO | 9661.469 | 17.05 | SAAO[a] |
| 9659.835 | 16.823 | CTIO | 9661.516 | 17.07 | SAAO[a] |
| 9659.842 | 16.832 | CTIO | 9661.570 | 17.07 | SAAO[a] |
| 9659.848 | 16.834 | CTIO | 9661.802 | 16.832 | CTIO |
| 9659.854 | 16.833 | CTIO | 9661.809 | 16.840 | CTIO |
| 9659.860 | 16.835 | CTIO | | | |

Note. — [a] From average of eight CCD images taken at the SAAO Sutherland observing station.

Table 2: Radial Velocities of 0513−69

| HJD (2440000+) | He II Peak (km s$^{-1}$) | Cross Corr. (km s$^{-1}$) | HJD (2440000+) | He II Peak (km s$^{-1}$) | Cross Corr. (km s$^{-1}$) |
|---|---|---|---|---|---|
| 8972.653 | 308 | 24 | 9666.778 | 288 | 1 |
| 8972.708 | 302 | 12 | 9666.846 | 291 | 8 |
| 8972.811 | 282 | −5 | 9666.859 | 292 | 5 |
| 8973.625 | 272 | −16 | 9667.652 | 287 | 2 |
| 9332.746 | 296 | 3 | 9667.665 | 282 | −2 |
| 9333.651 | 288 | −15 | 9667.747 | 283 | −5 |
| 9333.843 | 302 | −18 | 9667.824 | 282 | −8 |
| 9334.612 | 288 | −16 | 9667.840 | 276 | −10 |
| 9334.624 | 287 | −7 | 9668.689 | 264 | −16 |
| 9334.668 | 308 | 8 | 9668.702 | 276 | −7 |
| 9334.679 | 300 | −9 | 9668.760 | 266 | −25 |
| 9334.782 | 313 | 9 | 9668.771 | 264 | −21 |
| 9334.794 | 310 | 8 | 9669.697 | 282 | −2 |
| 9335.622 | 298 | 4 | 9669.714 | 276 | −8 |
| 9335.634 | 308 | 16 | 9669.859 | 276 | −3 |
| 9335.766 | 296 | 7 | 9670.634 | 282 | −4 |
| 9335.863 | 276 | ... | 9670.646 | 298 | 13 |
| 9336.570 | 320 | 7 | 9670.705 | 289 | 7 |
| 9336.631 | 308 | 14 | 9670.718 | 282 | 0 |
| 9336.851 | 296 | ... | 9670.824 | 280 | −3 |
| 9666.694 | 284 | −1 | 9670.836 | 274 | −10 |
| 9666.762 | 284 | 2 | | | |

Table 3: Average equivalent widths (Å)

| Wavelength (Å) | ID | 1992 | 1993 | 1994 |
|---|---|---|---|---|
| 4340 | H$\gamma$ | 4.8 | 1.2 | 3.3 |
| 4861 | H$\beta$ | 13.1 | 3.5 | 8.8 |
| 4200 | He II | 0.6 | 0.3 | 0.6 |
| 4541 | He II | 1.1 | 0.9 | 1.3 |
| 4686 | He II | 22. | 9.4 | 19. |
| 5411 | He II | ... | 2.1 | 2.9 |
| 4624 | S$^-$ | 1.1 | 0.6 | 0.7 |
| 4748 | S$^+$ | 0.8 | 0.5 | 0.7 |
| 4795 | S$^-$ | 0.4 | 0.3 | 0.3 |
| 4927 | S$^+$ | ... | 0.8 | 0.8 |

---

Table 4: Wavelengths (Å) and Velocities of shifted lines

| Year | He II 4686Å | | Hβ | |
|---|---|---|---|---|
| | S$^-$ | S$^+$ | S$^-$ | S$^+$ |
| 1992 | 4635.9 | 4748.8 | 4801.4 | ... |
| 1993 | 4627.7 | 4750.9 | 4798.8 | 4931.1 |
| 1994 low | 4628.1 | 4753.3 | 4798.7 | 4931.3 |
| 1994 high | 4628.3 | 4753.3 | 4801.0 | 4931.9 |
| 1993-4 mean | 4628.0 | 4752.5 | 4799.5 | 4931.4 |
| | | | | |
| mean RV (km s$^{-1}$) | −3974 | +3996 | −4060 | +4059 |

Fig. 1.— $V$ filter photometry of 0513−69 during November 1994. Filled symbols are data from individual CCD frames taken at CTIO while each open symbol represents the average of eight CCD images taken at the SAAO Sutherland observing station (see text).

Fig. 2.— Average spectra of 0513−69 and CAL 83 in 1994.

Fig. 3.— The He II to H$\beta$ region of 0513−69 in 1993 and 1994, showing the intensity variations of the emission lines and changes in line profiles.

Fig. 4.— Radial velocities of 0513−69 in 1994 Dec. measured from He II 4686Å peaks and from the entire spectrum through cross correlation of individual spectra (upper) against the mean.

Fig. 5.— Radial velocities derived by cross-correlation techniques of spectra of 0513−69 taken in 1992, 1993, and 1994, phased on P = 0.76 days. Velocities from the higher dispersion spectra in 1994 are indicated by larger points. The curve shows the fit to a circular orbit.

Fig. 6.— The relationship between possible stellar masses of 0513−69 for various values of the inclination, $i$, derived from the mass function, f(M) = $0.00011 M_\odot$, as discussed in the text. A main sequence companion would not fill its Roche lobe for any of the masses and inclinations shown on the diagram.

Fig. 7.— Details of the spectral line profiles. Upper: He II region, showing the weak emission lines and indicating possible ions which might contribute to the broad emission feature. Also marked are the displaced components 'S$^-$' and 'S$^+$'. The dashed line represents a symmetric profile, obtained by reflecting the red wing about the line center. Middle: H$\delta$ region, with major features identified. Lower: H$\beta$ region, showing the P Cygni absorption in the violet wing and the negatively displaced component 'S$^-$'.

Fig. 8.— The 4600Å – 4960Å region of 0513−69 in 1992, 1993, and 1994 (high and low resolution), showing the shifted lines (indicated by S$^-$ and S$^+$) on either side of He II 4686Å and H$\beta$. Note that the displacement of these components appears to be slightly smaller in 1992.